\documentclass[aps,twocolumn,superscriptaddress,prb,floatfix]{revtex4}
\usepackage{amssymb}
\usepackage{amsmath}
\usepackage{graphicx}

\def\Gq{e^2/h}
\def\ltheor{\xi}
\def\le{\xi_{\rm{fit}}}
\def\ls{\xi_s}
\def\SiOx{\mathrm{SiO}_2}
\def\Hethree{^3\mathrm{He}}

\def\den{n_{s}}
\def\Vbg{V_{\mathrm{bg}}}
\def\Voffset{V_{\mathrm{offset}}}

\def\be{\begin{equation}}
\def\ee{\end{equation}}

\begin{document}
\title{Quantum Hall conductance of two-terminal graphene devices}
\author{J. R. Williams}
\affiliation{School of Engineering and Applied Sciences, Harvard
University, Cambridge, MA 02138, USA}
\author{D. A. Abanin}
\affiliation{Department of Physics, Massachusetts Institute of
Technology, Cambridge, MA 02139, USA}
\author{L. DiCarlo}
\altaffiliation{Present address: Department of Applied Physics, Yale
University, New Haven, Connecticut 06520, USA}
\affiliation{Department of Physics, Harvard University, Cambridge,
MA 02138, USA}
\author{L. S. Levitov}
\affiliation{Department of Physics, Massachusetts Institute of
Technology, Cambridge, MA 02139, USA} \affiliation{Kavli Institute
for Theoretical Physics, University of California, Santa Barbara, CA
93106}
\author{C. M. Marcus}
\affiliation{Department of Physics, Harvard University, Cambridge, MA 02138, USA}

\date{\today}

\begin{abstract}
Measurement and theory of the two-terminal conductance of monolayer
and bilayer graphene in the quantum
Hall regime are compared. We examine features of
conductance as a function of gate voltage that allow
monolayer, bilayer, and gapped samples to be distinguished, including N-shaped distortions of quantum Hall plateaus and conductance peaks and dips at the charge neutrality point.
Generally good agreement is found between measurement and theory.
Possible origins of discrepancies are discussed.
\end{abstract}

\maketitle

\section{Introduction}

Graphene monolayers and bilayers are recently discovered
two-dimensional gapless semimetals. The Dirac spectrum of
excitations in monolayer graphene gives rise to a number of novel
transport properties, including anomalous quantized Hall
conductance with plateaus at $4(n+1/2)\,\Gq, \, n=0,\pm 1, \pm
2,...$ in multi-terminal samples.\cite{Novoselov05, Zhang05} Bilayer
graphene has a quadratic, electron-hole-symmetric excitation
spectrum, leading to quantized Hall conductance values $4n\,\Gq, \,
n=\pm 1,\pm 2,...$.\cite{Novoselov06, Geim07} Both monolayer and bilayer graphene have a zeroth Landau
level, located at the charge neutrality point (CNP), which is eightfold degenerate in bilayers and fourfold degenerate in monolayers. Other Landau levels are all fourfold degenerate in both
types of graphene.\cite{Gusynin05,McCann06} The novel transport signatures not
only reflect  this underlying band structure, but serve as
an experimental tool for identifying the number of layers and
characterizing sample quality.\cite{Geim07}

In recent work on graphene, two-terminal magnetoconductance has emerged
as one of the main tools of sample
characterization.\cite{Williams07, Ozyilmaz07, Heersche07} While a
two-terminal measurement is not as straightforward to interpret as
the corresponding multi-terminal measurement\cite{Beenakker91}, it is
 the simplest to perform and may be the only measurement possible,
for instance with very small samples. The presence of non-zero
longitudinal conductivity causes quantum Hall plateaus measured in a two-terminal configuration to not be as well quantized as in
multiprobe measurement.\cite{Geim07}  As discussed in detail below, plateaus
exhibit a characteristic N-shaped distortion arising from
the finite longitudinal conductivity that depends on device geometry.

In this Article, we systematically examine two-terminal conductance
in the QH regime for monolayer and bilayer graphene for a variety of
sample aspect ratios (Table I). We especially focus on the
features that can help to distinguish monolayer and bilayer graphene: the
conductance extrema in the N-shaped distortions of the quantum Hall plateaus and at the CNP.
We find that these features depend both on the sample aspect ratio and the
number of graphene layers. Results are compared to recent
theory\cite{Abanin08}, in which two-terminal conductance for
arbitrary shape is characterized by a single parameter $\ltheor$, the
effective device aspect ratio ($\ltheor = L/W$ for rectangular samples,
where $L$ is the length or distance between contacts, and $W$ is the
device width).

\begin{table}[ht]
\caption{Measured two-terminal graphene devices.}
\centering
\begin{tabular}{c c c c c}
\hline\hline Sample & Layers (Inferred) & ($L$, $W$) [${\rm \mu m}$] &
$\ls$ & $\le$ \\ [0.5ex]
\hline
A1 & Monolayer & (1.3, 1.8) & 0.7 & 1.7\\
A2 & Monolayer & (0.4, 2.0) & 0.2 & 0.2\\
B1 & Bilayer & (2.5, 1.0) & 2.5 & 0.8\\
B2 & Bilayer & (0.3, 1.8) & 0.2 & 0.3\\
C & Monolayer & Asymmetric & 0.9\footnotemark[1] & 0.9 \\
[1ex] \hline\hline
\end{tabular}
 \footnotetext[1]{effective aspect ratio, see Sect. IV.}
\label{table:nonlin}
\end{table}

In Ref.\,[\onlinecite{Abanin08}], the positions of
conductance extrema on the distorted plateaus were linked to
incompressible densities. Here we find that this relation can be
used to distinguish monolayer and bilayer graphene devices even
when the distortions of the plateaus are strong. The analysis of
rectangular two-terminal samples is extended to a sample with
asymmetric contacts, extracting an effective sample aspect ratio via
conformal mapping. Best-fit values of the aspect ratio, $\le$,
obtained by fitting the theory to the experimental data, are
compared to the measured sample aspect ratio, $\ls$. Agreement is
generally good, but not uniformly so. We speculate on possible
causes of these discrepancies, including inhomogeneous contact
resistance, electron and hole puddles, and contributions of
transport along \emph{p-n} interfaces.

\subsection{Qualitative Discussion}

Representative theoretical plots of two-terminal conductance for
monolayer, bilayer, and gapped bilayer graphene as a function of
filling factor, $\nu$, are shown in Fig.~1. For both monolayers and
bilayers, the absence of an energy gap between the conduction and
valence bands gives rise to a zero energy Landau level
(LL)\cite{Gusynin05}, which can either increase or decrease the
two-terminal conductance around the charge neutrality point,
depending on the aspect ratio of the sample. The eightfold
degeneracy of the zero-energy LL in bilayer graphene\cite{McCann06}
enhances the size of this feature relative to monolayer graphene.

A gap in the spectrum of bilayer graphene opens when the on-site
energy in one layer differs from the on-site energy in the
other.\cite{McCann06a}  This may result, for instance, from
asymmetric chemical doping\cite{Castro07} or electrostatic
gating.\cite{Oostinga08} The gap splits the zero-energy LL,
suppressing conductance at the CNP. The qualitative effect of a
gap in the bilayer spectrum can be seen in Fig.~1 by comparing the
gapped case [Fig.~1(c)], which always has a zero of conductance at
$\nu=0$, to the gapless cases [Figs.~1(a,b)], which has a non-zero
value of conductance at  $\nu=0$.

Also illustrated in Fig.~1 is how the aspect ratio of the sample
affects the two-terminal conductance near quantum Hall plateaus for
all three spectrum types. Finite longitudinal conductivity leads to
N-shaped distortions of the plateaus,\cite{Abanin08} which are of
opposite signs for aspect ratios $\ltheor <1$ and $\ltheor >1$. Note, however, that the extrema of conductance---minima for $\ltheor <1$ and maxima for
$\ltheor >1$---are aligned with the plateaus centers, which coincide with the incompressible density values (different for monolayers and bilayers). Distorted plateaus thus remain useful for characterizing the number of layers and density.

The back-gate dependence of conductance for the
five samples reported are most similar to those in
Figs.~1(a,b), indicating that these samples are single layers and
gapless bilayers only (see Table 1). We use the model of
Ref.\,[\onlinecite{Abanin08}] to fit the conductance data treating
the aspect ratio as a fit parameter. In doing so, our
presumption is that the visible dimensions of the sample may not
reflect the actual pattern of current flow. Since the conductance
problem for a sample of any shape can be reduced to that of an
effective rectangle via a conformal
mapping \cite{Wick54,Jensen72,Rendell81}, which depends on the sample
shape but \emph{not} on the conductivity tensor, the rectangular
geometry is universal for two-terminal conductance. Thus the model
of a conducting rectangle with an unspecified aspect ratio is
suitable for describing systems in which current pattern is not
precisely known.

\begin{figure}[t!]
\center
\includegraphics[width=2.8in]{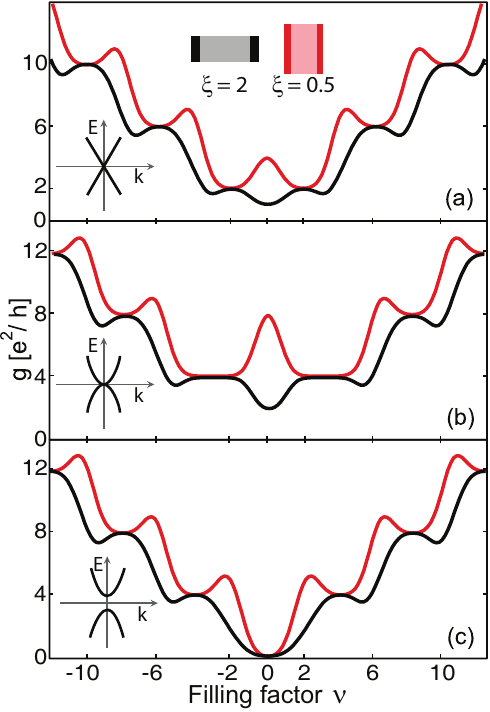}
\caption{\footnotesize{ (color online) Theoretical two-terminal QH
conductance $g$ as a function of filling factor $\nu$
(Ref.\,[\onlinecite{Abanin08}]) shown for (a) single-layer graphene
(b) bilayer graphene (c) gapped bilayer graphene for effective
aspect ratios $\ltheor=L/W=2$ (black) and 0.5 (red). Finite longitudinal
conductivity due to the states in the middle of each Landau level
distorts the plateaus into N-shaped structures, which are of
opposite sign  for $\ltheor<1$ and $\ltheor>1$. Local extrema of $g$ at
filling factors $\nu=\pm2, \pm6, \pm10...$ for single layers and at
$\nu=\pm4, \pm8, \pm12...$ for bilayers are either all maxima
($\ltheor<1$) or all minima ($\ltheor>1$). For gapless monolayer and bilayer
samples (a,b), $g(\nu=0)$ is a minimum for $\ltheor<1$ and maximum for
$\ltheor>1$; for the gapped bilayer (c) $g$ vanishes at $\nu=0$ for all
$\ltheor$.
 }} \label{fig0}
\end{figure}

\subsection{Sample Fabrication and Measurement}

Graphene devices were fabricated by mechanically exfoliating highly
oriented pyrolytic graphite\cite{Novoselov04} onto a $n^{++}$ Si
wafer capped with $\sim 300\,{\rm nm}$ of $\SiOx$. Potential single
and bilayer graphene flakes were identified by optical microscopy.
Source and drain contacts, defined by electron beam lithography,
were deposited by thermally evaporating $5/40\,{\rm nm}$ of Ti/Au.
The aspect ratio, $\ls$, of each sample was measured using either optical or
scanning electron microscopy.

Devices were measured in a $\Hethree$ refrigerator allowing dc
transport measurements in a magnetic field $|B|<$ $8\,\mathrm{T}$
perpendicular to the graphene plane. Unless otherwise noted, all
measurements were taken at  base temperature, $T\sim250\mathrm{mK}$.
Differential conductance $g=dI/dV$, where $I$ is the current and $V$
the source-drain voltage, was measured using a current bias ($I$
chosen to keep $eV<k_BT$) and standard lock-in technique at a
frequency of 93Hz. All samples show $B=0$ characteristics of
high-quality single-layer and bilayer graphene\cite{Novoselov05,
Zhang05}: a CNP positioned at back-gate voltage $\Vbg\sim0$ and a
large change in $g$ (in excess of $20\,\Gq$) over the $\Vbg$ range
of $\pm 40\,{\rm V}$.

\section{Monolayer Samples}
Figure 2(a) shows the two-terminal conductance $g(\Vbg)$ for sample A1
($\ls =0.7$) at B= 8 T (black trace). Plateaus are seen at
$\nu=\pm 2$ near---but not equal to---$2\,\Gq$, with values of
$\sim2.3(2.7)\,\Gq$ on the electron (hole) side of
the CNP. At the CNP ($\Vbg\sim2.3\mathrm{V}$), $g$
departs from the quantized values, dropping to a minimum of
$\sim1.4\Gq$. At higher densities, the conductance exhibits a series
of maxima with values slightly above 6, 10, $14\,\Gq$. Maxima on the
hole side consistently have slightly higher values, a feature
observed in all the samples measured. The inset of Fig.~2(a) shows
$g$ in the QH regime as a function of $\Vbg$ and $B$. Dashed black
lines indicate the filling factors $\nu=\den h/eB$ (where $\den$ is
the carrier density) of $-6$, $-10$, and $-14$ and lines align with the
local {\it maxima} of $g(\Vbg,B)$. $\Vbg$ was converted to $\den$
using a parallel plate capacitance model\cite{Novoselov04}, giving
$\den=\alpha (\Vbg+\Voffset)$ with
$\alpha=6.7\times10^{10}\mathrm{cm^{-2}}\mathrm{V^{-1}}$ and
$\Voffset=2\mathrm{V}$.

\begin{figure}[t!]
\center
\includegraphics[width=3.05in]{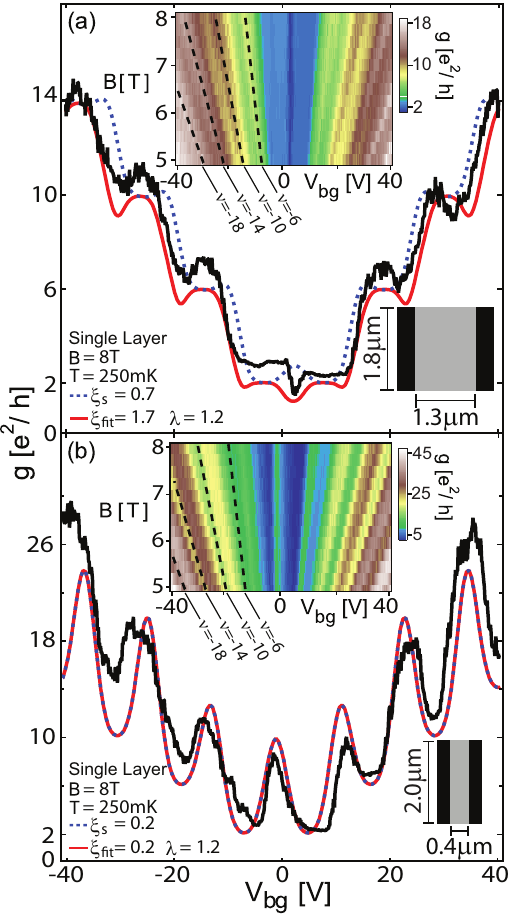}
\caption{\footnotesize{ (color online) (a) Inset: Conductance $g$ in the
quantum Hall regime as a function of $B$ and $\Vbg$ at
$\mathrm{T}=250\,{\rm mK}$ for sample A1. Black dashed lines
correspond to filling factors of $\nu =-6, -10, -14, -18$ and align
with the local \emph{maxima} of conductance. Main: (black)
Horizontal cut of inset giving $g(\Vbg)$ at $B=8\,{\rm T}$ and
calculated $g$ for the best-fit equivalent aspect ratio $\le=1.7$
(solid red curve) and the actual sample aspect ratio $\ls=0.7$
(dashed blue curve) using Landau level broadening parameter
$\lambda=1.2$. (b) Inset: Conductance $g$ in the quantum Hall regime
as a function of $B$ and $\Vbg$ at $\mathrm{T}=250\,{\rm mK}$ for
sample A2. Black dashed lines correspond to filling factors of
$\nu=-6, -10, -14, -18$ and align with the local \emph{minima} of
conductance. Main: (black) Horizontal cut of inset giving $g(\Vbg)$
at $B=8\,{\rm T}$ and calculated $g$ for $\le=0.2$ (solid red curve)
and $\ls=0.2$ (dashed blue curve) ($\lambda=1.2$, the same as sample
A1).
 }} \label{fig1}
\end{figure}

Measured $g(\Vbg)$ [black curve in Fig.~2(b)] for sample A2
($\ls=0.2$), made using the same graphene flake as A1, shows
distinctive differences from the measured $g(\Vbg)$ of sample A1. In
particular, at the CNP ($\Vbg=-1.5\mathrm{V}$), $g$ exhibits a sharp
peak with a maximal value of $\sim 8.8\,\Gq$. Away from the CNP, the
conductance has maxima which are much stronger than those of sample
A1. The inset of Fig.~2(b) shows $g(\Vbg,B)$. For this sample, the
dashed lines representing the incompressible filling factors $\pm 6,
\pm 10, \pm 14$ now align with the {\it minima} in $g$. Here we used
the $\Vbg$ to $\den$ conversion factors of
$\alpha=6.7\times10^{10}\mathrm{cm^{-2}}\mathrm{V^{-1}}$ (the same
as for sample A1) and $\Voffset=-1.1\mathrm{V}$.

The observed features in $g$  for samples A1 and A2 can be  compared
to theory\cite{Abanin08} for two-terminal quantum Hall conductance,
which uses a model of a conducting rectangle $L\times W$ with a
spatially uniform conductivity. The filling factor dependence of the
conductivity tensor is obtained using the semicircle relation for
quantum Hall systems, derived in Ref.\,[\onlinecite{Dykhne94}],
which is applied independently for each Landau level. Landau level
broadening due to disorder is included in the theory as a gaussian
broadening $e^{-\lambda(\nu-\nu_n)^2}$, where $\nu_n$ is the center
of the LL and $\lambda$ is a fitting parameter. The total
conductivity tensor is taken to be a sum of the contributions of
individual Landau levels.  The current-density distribution for a
rectangular sample with an arbitrary aspect ratio is found
analytically by conformal mapping (see
Refs.\,[\onlinecite{Wick54,Jensen72,Rendell81}]). The current
density is then integrated numerically along suitably chosen
contours to evaluate total current and voltage drop, from which
$g=I/V$ is obtained.

Along with the experimental traces, Figs.~2(a,b) also show the
theoretical curves for the best-fit (solid red trace) and the actual
sample aspect (dashed blue trace) ratios.  For sample A1, the
best-fit aspect ratio, $\le =1.7$, differs considerably from the
measured value, $\ls =0.7$. For sample A1, the best fit gives the
Landau level broadening parameter $\lambda=1.2$. This theoretical
curve ($\le=1.7$) reproduces the essential features of the data:
local maxima align with the filling factors $\pm 2,\pm 6, \pm 10,$
..., and $g$ exhibits a dip at the CNP.

The alignment of conductance minima with densities corresponding to
the integer filling factors as well as a peak at the CNP observed
for sample A2 are consistent with theoretical predictions for a
short, wide monolayer graphene sample. As illustrated in Fig.~2(b),
the best-fit aspect ratio $\le=0.2$ agrees well with the measured
$\ls$ for sample A2.

We observe that the size of peaks and dips in Fig.~2(a,b) increases
for higher LL. In contrast, theory\cite{Abanin08} predicts that
peaks and dips at $|\nu|>0$ LLs are all roughly the same. This
discrepancy may reflect the inapplicability of the two-phase model
approach of Ref.\,[\onlinecite{Dykhne94}], which underlies the
semicircle law obtained in this work, to higher LLs. Indeed, because
for Dirac particles the spacing between LLs decreases at higher
energies as an inverse square root of the level number, one may
expect mixing between non-nearest LLs to increase at high energies.
Such mixing can lead to the longitudinal conductivity values in
excess of those of Ref.\,[\onlinecite{Dykhne94}], which only
considers mixing between nearest LLs (see the discussion in
Ref.\,[\onlinecite{Burgess07}]).

To take these effects into account, we extend the model of
Ref.\,[\onlinecite{Abanin08}] by assuming that the contribution of
the $n^{\mathrm{th}}$ LL to the conductivity tensor in monolayer
graphene is described by a modified semicircle (``elliptic") law,
\be\label{eq:modified} \delta_n\sigma_{xx}^2+A_n^2
(\delta_n\sigma_{xy}-\sigma_{xy,n}^0)(\delta_n
\sigma_{xy}-\sigma_{xy,n'}^0)=0, \ee
where $\delta_n \sigma_{xx}$
and $\delta_n \sigma_{xy}$ are the effective longitudinal and Hall
conductivities, $\sigma_{xy,n}^0$ and $\sigma_{xy,n'}^0$ are the
quantized Hall conductivities at the neighboring plateaus. Here $n$
and $n'$ are neighboring LL indices, related by $n'=n+1$ (except the
doubly degenerate $\nu$ =0 LL for the bilayer, in which case $n=-1$
and $n'=1$). The $A_n$ account for departures from the semicircle
law. We take $A_n\approx 1$ for $n=0,\pm 1$, and $A_n\approx 2$ for
other LLs, consistent with previous observations.\cite{Burgess07}

\begin{figure}[t!]
\center
\includegraphics[width=3.25in]{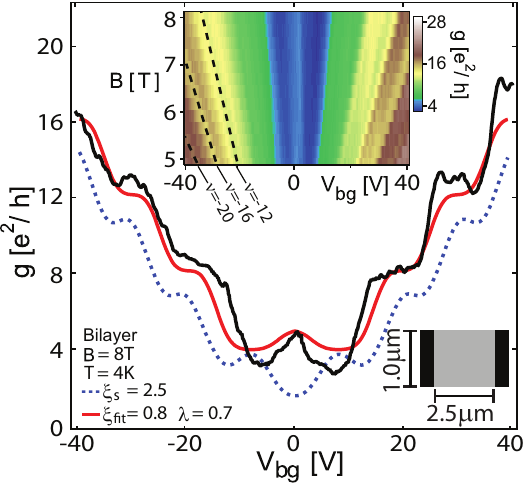}
\caption{\footnotesize{ (color online) Inset: Measured  $g$ of sample B1 as
a function of $B$ and $\Vbg$ at $T= 4 \mathrm{K}$. Black dashed
lines, corresponding to $\nu= -12, -16, -20$, align with local
\emph{minima} of $g$. No minima are observed at $\nu = 8$ for
$5\mathrm{T}<B<8\mathrm{T}$. Main: Horizontal cut of inset at $B =
8\mathrm{T}$ (black), and calculated $g$ using $\lambda = 0.7$ for
$\ls$ = 2.5 (dashed blue curve) and $\le = 0.8$ (solid red curve).
}}
 \label{fig2}
\end{figure}

\section{Bilayer Samples}

The black curve in Fig.~3 shows measured $g(\Vbg)$ for sample B1
($\ls=2.5$) at $B=8$T and $T=4$K.  This sample has two features
indicating that it is a bilayer sample: plateaus in conductance
appearing near 4, 8, 12 and $16\,\Gq$, and a conductance maximum at
the CNP whose relative size is much larger then those at higher LLs. The
conductance values at the plateaus $\nu=\pm 4$ here are lower than
the expected $4\,\Gq$ for a bilayer sample, falling to
$2.7(3.1)\,\Gq$ on the electron (hole) side of the CNP. The peak
value in conductance at $\nu=0$ ($\Vbg=0.5\,{\rm V}$) is $5\,\Gq$.
At higher filling factors, the plateaus exhibit two different
behaviors, showing a flat plateau at $\nu=8$ and a plateau
followed by a dip at $\nu=12$. The small dips align with the filling
factors $\nu=-12, -16, -20$ for $5\, {\rm T}<B< 8\, {\rm T}$ (see
inset of Fig.~3), using
$\alpha=7.2\times10^{10}\mathrm{cm^{-2}}\mathrm{V^{-1}}$ and
$\Voffset=0.5\mathrm{V}$.

Theoretical $g$ curves for aspect ratios $\ls=2.5$ (dashed blue
curve) and $\le=0.8$ (solid red curve) are shown in Fig.~3.
Theoretical $g(\Vbg)$ curves for these two aspect ratios are similar
at high density, but differ for $\nu=0$: the curve for $\ls=2.5$ has
a dip in conductance at the CNP while the best-fit curve ($\le=0.8$)
has a peak, similar to the experimental curve. The curve for
$\le=0.8$ also agrees better with experiment at higher densities.

\begin{figure}[t!]
\center
\includegraphics[width=3.2in]{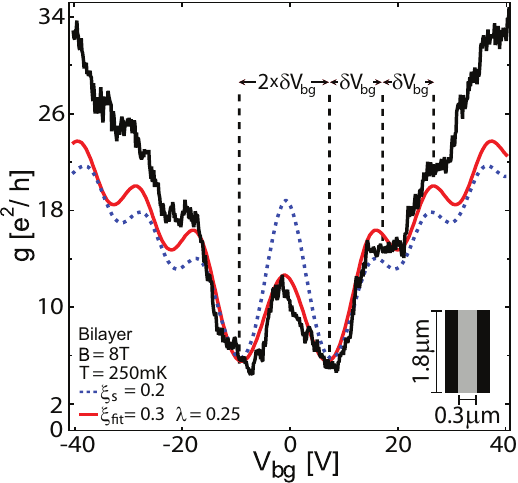}
\caption{\footnotesize{ (color online) Measured $g(\Vbg)$ for sample B2
(black) and the calculated $g$ using $\lambda=0.25$ for $\ls=0.2$
(dashed blue trace) and $\le=0.3$ (solid red trace). Two key
features in the curve suggest this sample is a gapless bilayer,
namely, a pronounced peak in $g$ near the CNP, and the larger
spacing between the two minima straddling the CNP compared to the
spacing $\delta \Vbg \sim 9.5$ between other consecutive minima.
 }} \label{fig4}
\end{figure}

In some cases the quantized conductance values are found to be quite
different from the expected quantized values, as demonstrated in
Fig.~4 for sample B2 ($\ls=0.2$). In this sample, $g$ reaches a
maximum of $13.5\,\Gq$ at the CNP, with adjacent minima of $5\,\Gq$.
Away from the CNP, conductance plateaus appear at values of
$\sim16\,\Gq$ and $23\,\Gq$, neither of which are near expected
values  for monolayer or bilayer graphene. Since there are no strong
peaks or dips in $g$ away from charge neutrality, as is expected for
a device with a $\ls\ll1$, it is difficult to determine the number
of layers from the location of the conductance extrema.  There are
two conductance features, however, that suggest the sample is
gapless bilayer graphene. First, the peak at $\nu=0$ is much more
pronounced than any other peak in the conductance. Second, the
spacing in $\Vbg$ between the two lowest LLs is twice as large as
the spacing between any other two successive LLs (in Fig.~4, $\delta
\Vbg\sim 9.5\,{\rm V}$). Both features arise in bilayers as a result
of the zero-energy LL being eightfold degenerate, twice as much as
all other bilayer LLs and the zero-energy LL in single layer
graphene.\cite{McCann06} The theoretical prediction, using $\le
=0.3$ (solid red line) and $\ls=0.2$ (dashed blue line), for the
bilayer sample B2 are shown in Fig.~4.

\section{Non-rectangular samples}

In this section we extend the comparison of theory and experiment to a
non-rectangular device, sample C, shown schematically in
the inset of Fig.~5. The measured two-terminal conductance of
sample C (black curve in Fig.~5) has properties very similar to
those expected for a square monolayer sample: around the CNP the
conductance is nearly flat with value close to $2\,\Gq$,
monotonically increasing on the electron and hole sides at filling
factors $|\nu|> 2$.

\begin{figure}[t!]
\center
\includegraphics[width=3in]{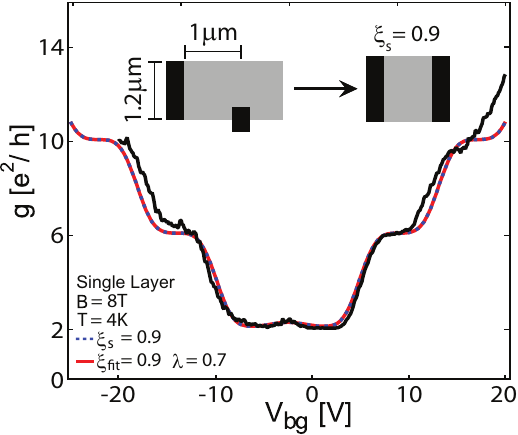}
\caption{\footnotesize{ (color online) Measured $g(\Vbg)$ for sample C
(black) and calculated conductance (solid red curve) for $\le=0.9$
($\lambda=0.7$). The asymmetric contacts of this sample can be
conformally mapped onto a rectangle, producing a device aspect ratio
of $\ls=0.9$ (dashed red curve, directly under solid red curve). }}
\label{fig5}
\end{figure}

Theoretical curve shown in Fig.\,5 is obtained from the conducting
rectangle model using a best-fit effective aspect ratio $\le=0.9$
and the LL broadening parameter $\lambda=0.7$. This choice of
parameters yields particularly good agreement for $|\nu|\leq 6$. At
higher fillings, the plateaus are washed out, suggesting that the LL
broadening is stronger for LLs $|n|\geq 2$. It is interesting to
compare the best-fit value $\le$ to an effective aspect ratio,
obtained from conformal mapping of sample C to a rectangle. As
discussed below, this conformal mapping can be constructed directly,
owing to the relatively simple geometry of sample C. The effective
aspect ratio obtained in this way is $\ls\approx 0.9$, which is
consistent with the best-fit value.

\begin{figure}[t!]
\center
\includegraphics[width=2.0in]{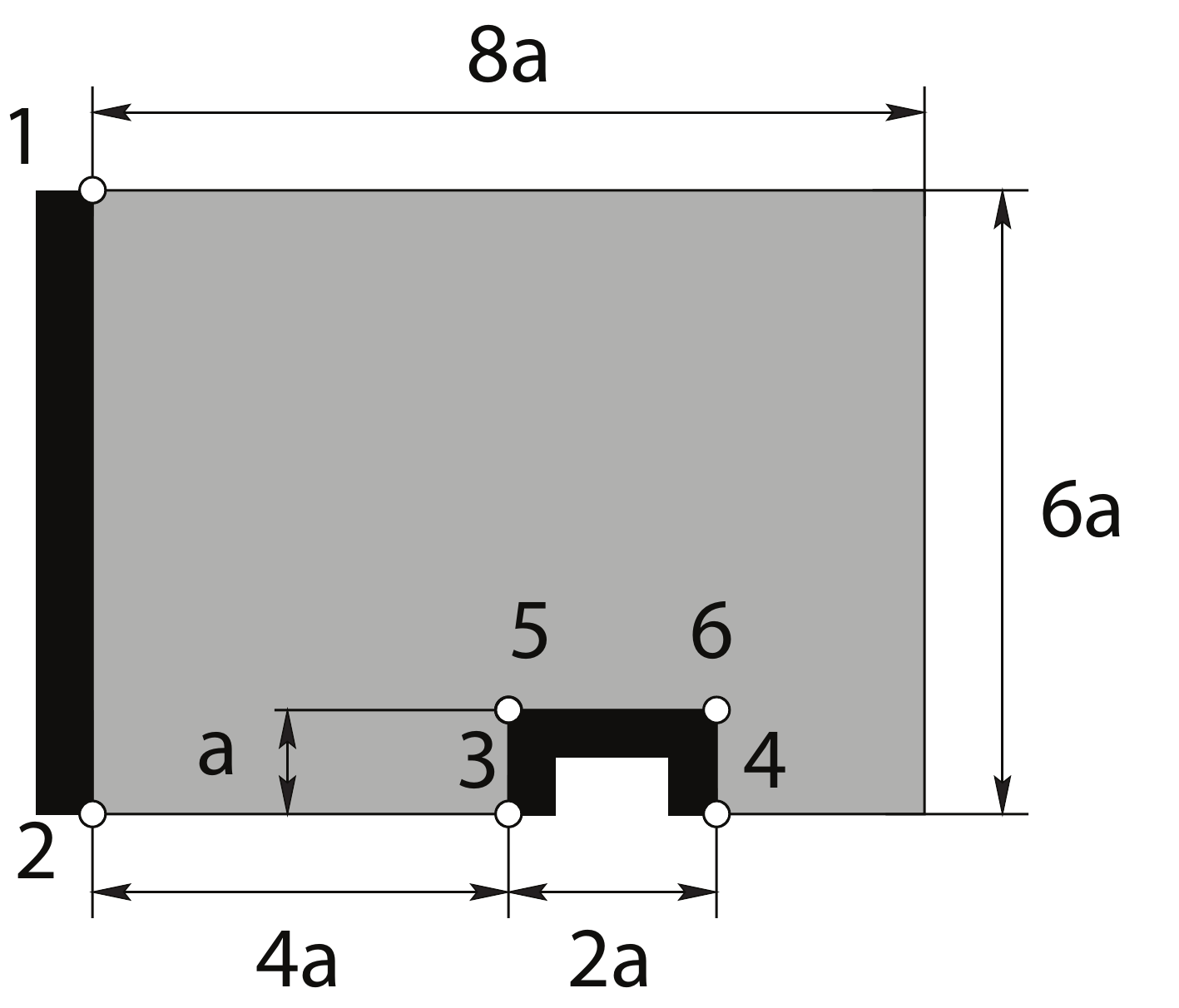}
\caption{ (color online) A polygon representing sample C (see Fig.~5).
Black regions correspond to contacts, length scale $a=200\,{\rm
nm}.$}
 \label{fig6}
\end{figure}

Before we proceed to construct the conformal mapping we note that
the geometry of sample C, pictured in Fig.~6, is that of a polygon.
In principle, any polygon can be mapped onto the upper half-plane by
inverting a Schwarz-Christoffel mapping.\cite{Driscoll02} However,
since this mapping is defined by a contour integral, the inverse
mapping can only be found numerically. In order to circumvent this
difficulty, two approximations are employed below, allowing the
desired mapping to be constructed as a composition of a few simple
mappings.

The steps involved in this construction are illustrated in Fig.~7.
First, the rectangular shape in Fig.\,6 is replaced by a
semi-infinite strip shown in Fig.~7(a). This approximation should
not significantly affect the conductance, as the current flows
mostly in the region between contacts 1-2 and 3-4. Without loss of
generality we set the length scale $a=1$.

\begin{figure}[t!]
\center
\includegraphics[width=3.25in]{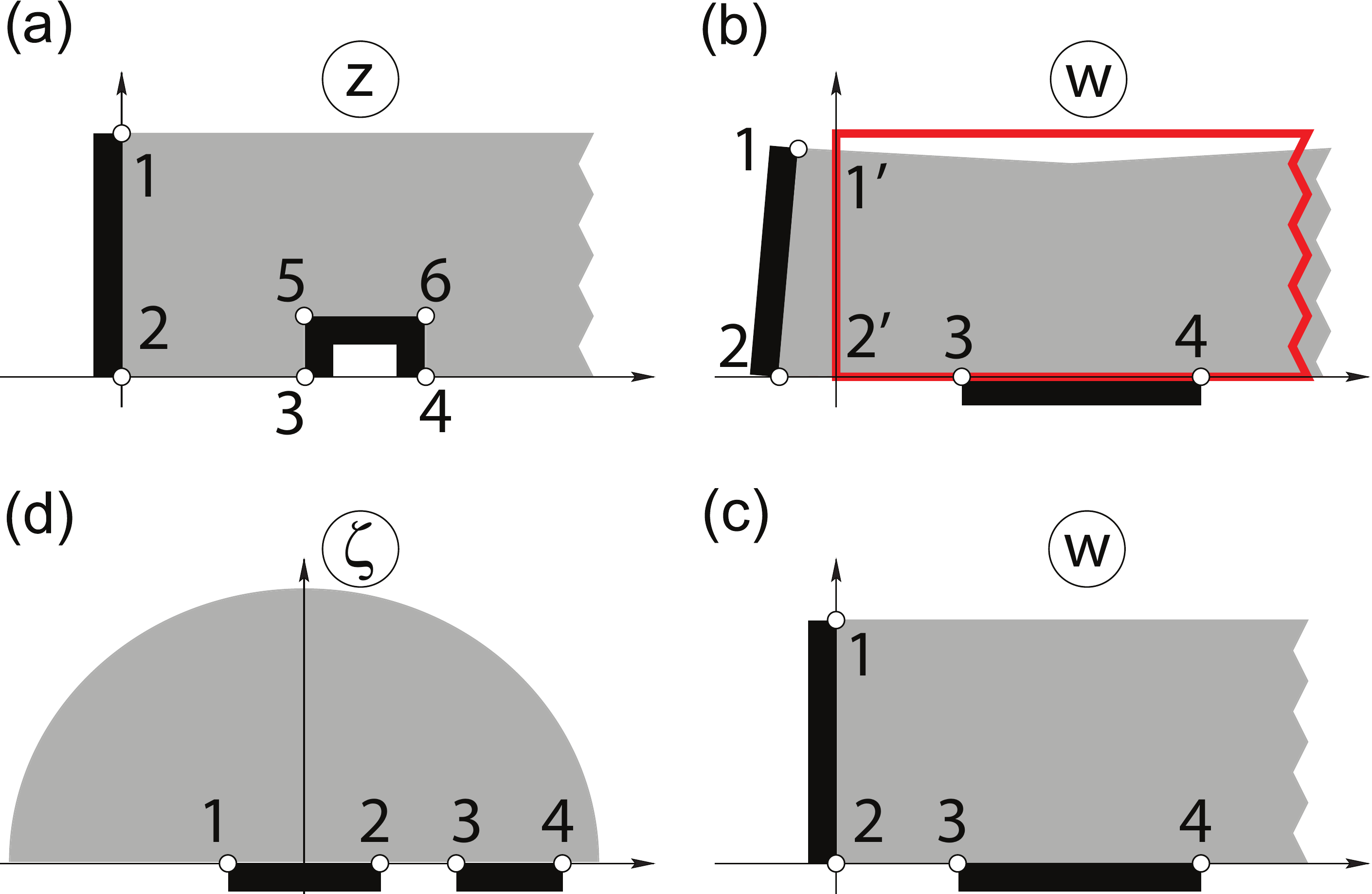}
\caption{ (color online) Three steps used to map the polygon in Fig.~6
(sample C) onto the upper half-plane (schematic). First, the
rectangle in Fig.~6 is replaced by a half-infinite strip, extending
indefinitely to the right (a). Next, we map the domain shown in (a)
onto a rectangle with contact 3-5-6-4 straightened out (b). Under
this mapping, the sample is slightly distorted, as indicated by the
grey polygon in (b). Because the deviation of the grey polygon
boundary from the original sample boundary (red line in (b)) is
fairly small, it can be neglected, giving a half-infinite strip (c).
Finally, the domain (c) is  mapped onto the upper half-plane (d),
which allows to find the cross ratio $\Delta_{1234}$,
Eq.~(\ref{eq:crossratio}), and evaluate the effective aspect ratio,
Eq.~(\ref{eq:xi_eff}).}
 \label{fig7}
\end{figure}

Our next step is to straighten out the contact 3-5-6-4. For that,
let us consider an auxiliary mapping that maps the upper $\tilde{w}$
plane onto the upper $\tilde{z}$ plane with a removed
rectangle~\cite{ConformalMapping}:
\be\label{eq:map1} \tilde{z}-iA=\int_0^{\tilde{w}} \left(
\frac{\xi^2-1}{\xi^2-2} \right)^{1/2}\, d\xi. \ee
We choose the parameter $A$ to be equal
\be\label{eq:A} A= \int_0^{1}   \left(
\frac{\xi^2-1}{\xi^2-2} \right)^{1/2}\, d\xi\approx 0.60
,
\ee
so that the removed rectangle has vertices
\be\label{C} \tilde{z}_{3,4}=\pm
A, \,\, \tilde{z}_{5,6}=\pm A+iA.
\ee
These points correspond to the points $\tilde{w}_{3,4}=\pm
\sqrt{2}$, $\tilde{w}_{5,6}=\pm 1$ in the $\tilde{w}$ plane. The
value of $A$ ensures that the edge of the sample on the $x$ axis
remains on the $x$ axis under the mapping (\ref{eq:map1}). The
distance between points $\tilde{z}_3$ and $\tilde{z}_5$ plane equals
$A$, as follows from Eq.~(\ref{eq:map1}) and the identity
\be\label{eq:equality} \int_1^{\sqrt{2}} \left|
\frac{\xi^2-1}{\xi^2-2} \right|^{1/2}\, d\xi = \int_0^{1} \left(
\frac{\xi^2-1}{\xi^2-2} \right)^{1/2}\, d\xi, \ee
which can be proved by making the change of variables,
$\xi=\sqrt{2-x}$ in the integral in the left-hand side of
Eq.~(\ref{eq:equality}), and $\xi=\sqrt{x}$ in the integral in the
right-hand side of Eq.~(\ref{eq:equality}).

The removed rectangle has aspect ratio equal to 2, the same as that for
the contact 3-5-6-4, however, their dimensions differ by a factor
of $A$. Scaling and shifting both $\tilde{z}$ in $\tilde{w}$,
\be\label{eq:ztildez} \tilde{z}=A(z-5), \,\, \tilde{w}=A(w-5),
\ee
we obtain the required mapping which straightens out the contact
3-5-6-4.

The second approximation is necessary because the mapping
(\ref{eq:map1}),~(\ref{eq:ztildez}), while straightening the
segments 3-5-6-4, distorts the rest of the boundary.  We notice,
however, that sufficiently far from the contact 3-5-6-4 the mapping
(\ref{eq:map1}) is close to the identity:
\be\label{eq:z=w} z(w\gg1)=w+O(1/w) ,\quad |z-5|\gg 1 . \ee
This property and the relatively small size of the segments 3-5-6-4
compared to the strip width guarantees that the distortion is small.
This is shown schematically in Fig.~7(b), where the curved grey
polygon represents the actual image of the sample, with the
deviation of its boundary from the strip of the same asymptotic
width (shown in red and exaggerated for clarity). The deviation is
indeed small: by investigating the mapping
(\ref{eq:map1}),~(\ref{eq:ztildez}) numerically we found that the
boundary is displaced the most at the point 2 which is shifted by
approximately $0.3$ away from its original position $2'$ along the
real axis. This is small compared to the sample width, equal to 6,
which allows us to neglect the displacement of the boundary. Thus we
assume that the mapping (\ref{eq:map1}),~(\ref{eq:ztildez})
transforms sample C into the semi-infinite strip shown in Fig.~7(c).

After this approximation is made,  it is straightforward to
transform the semi-infinite strip in Fig.~7(c) into the upper
half-plane, which can be done by the following mapping,
\be\label{eq:map2}
\zeta={\rm cosh} \frac{\pi w}{6}.
\ee
In the $\zeta$ plane, the contacts are mapped on the real axis, with
the end points 1, 2, 3 and 4  mapped to $\zeta_1=-1$, $\zeta_2=1$,
$\zeta_3\approx 2.11$, $\zeta_4\approx 23.57$. From these values,
following the procedure described in
Ref.\,[\onlinecite{Abanin08}] (Appendix), we compute the cross ratio
\be\label{eq:crossratio}
\Delta_{1234}=\frac{(\zeta_1-\zeta_4)(\zeta_3-\zeta_2)}{(\zeta_1-\zeta_2)(\zeta_3-\zeta_4)}\approx
-0.64,
\ee
and then obtain the aspect ratio from the
relations
\be\label{eq:xi_eff}
\ls=\frac{L}{W}=\frac{K(k')}{2K(k)} ;\quad
\Delta_{1234}=(1-k^2)/2k ,
\ee
where $K(k)$ is the complete elliptic integral of the first kind,
and $k'=(1-k^2)^{1/2}$. This procedure yields the value $\ls=0.9$,
identical to that found from the best fit to a conducting rectangle
model (see Fig.\,5).

\section{Summary and Discussion}

In summary, we have studied the effect of geometry on the
conductance of two-terminal graphene devices in the QH regime, comparing experiment and theory. The
quantized QH plateaus typically exhibit conductance extrema that are
stronger for wide, short samples. For non-rectangular samples, the
equivalent rectangle approach appears works well.

Theoretically, for short, wide samples ($\le<1$) the
two-terminal conductance a conductance minimum is expected at filling factors where
plateaus exist in multiterminal devices, while for long,
narrow samples ($\le>1$), a conductance  maximum is expected at these filling factors. Along with the behavior at the CNP, these
signatures provide a clear way to identify the number of layers in
the sample even when the quantization is weak or absent.

We find in the five samples measured that
conductance as a function of gate voltage is well described by theory, allowing for a phenomenological Landau Level broadening, and treating the aspect ratio as a fit parameter. In some samples, however, the best fit aspect ratio differs considerably from the measured aspect ratio of the sample.

What might lead to the discrepancy between some of the measured and
fit aspect ratios? One source of discrepancy could be that only part
of the contact actually injects current. It might also be that the
contacts locally dope the graphene, causing the aspect ratio to
appear smaller. This latter scenario, however, would require that
the doping penetrates $\sim 500\,{\rm nm}$ into the graphene, which
is $\sim 2$ orders of magnitude larger than
expected.\cite{Giovannetti07} Another, more interesting possibility
could be that the picture of an effective medium characterized by
local conduction, on which the argument leading up to the
semi-circle relation\cite{Dykhne94} is based, may not hold. This
might arise, for instance, from large density fluctuations, giving
rise to  electron and hole puddles\cite{Yacoby07} forming a network of
\emph{p-n} interfaces along which conduction occurs. In this case,
the effect of the back gate is to alter the percolation properties
of this \emph{p-n} network.  Transport mediated by such states would
almost certainly change the conventional picture of local
conduction. Further studies are required to clarify the physical
mechanism responsible for the observed behavior.

{\it Acknowledgement.} Research supported in part by INDEX, an NRI
Center, the Harvard NSEC, and the Harvard Center for Nanoscale
Systems (CNS), a member of the National Nanotechnology
Infrastructure Network (NNIN), which is supported by the National
Science Foundation under NSF award no. ECS-0335765. We thank
Pablo~Jarillo-Herrero for helpful discussions.


\begin{thebibliography}{10}

\bibitem{Novoselov05}
K. S. Novoselov \textit{et al.}, Nature \textbf{438}, 197 (2005).

\bibitem{Zhang05}
Y. Zhang \textit{et al.}, Nature \textbf{438}, 201 (2005).

\bibitem{Novoselov06}
K. S. Novoselov \textit{et al.}, Nat. Phys. \textbf{2}, 177
  (2006).

\bibitem{Geim07}
A. K. Geim and K. S. Novoselov, Nat. Mater. \textbf{6}, 183
(2007).

\bibitem{Gusynin05}
V.~P.~Gusynin and S.~G.~Sharapov, Phys. Rev. Lett. \textbf{95},
146801 (2005).

\bibitem{McCann06}
E. McCann and V. I. Falko, Phys. Rev. Lett. \textbf{96}, 086805
(2006).

\bibitem{Heersche07}
H. B. Heersche \textit{et al.}, \textit{Nature} \textbf{446}, 56
(2007).

\bibitem{Williams07}
J. R. Williams, L. DiCarlo and C. M. Marcus, Science \textbf{317},
638 (2007).

\bibitem{Ozyilmaz07}
B. \"{O}zyilmaz \textit{et al.}, Phys. Rev. Lett. \textbf{99},
166804 (2007).


\bibitem{Beenakker91}
C.~W.~J.~Beenakker and H.~van~Houten, \emph{Solid State Physics},
edited by H.~Ehrenreich and D.~Turnbull (Academic, New York, 1991),
Vol. 44, pg. 1.


\bibitem{Abanin08}
D. A. Abanin and L. S. Levitov, Phys. Rev. B \textbf{78}, 035416
(2008).

\bibitem{McCann06a} E. McCann,  Phys. Rev. B {\bf 74}, 161403(R) (2006).

\bibitem{Castro07}
E.~V.~Castro \textit{et al.}, Phys. Rev. Lett. {\bf 99}, 216802
(2007).


\bibitem{Oostinga08}
J.~B.~Oostinga \textit{et al.}, Nat. Mater. \textbf{7}, 151 (2007).

\bibitem{Wick54} R. F. Wick, J. Appl. Phys. \textbf{25}, 741 (1954).

\bibitem{Jensen72} H. H. Jensen and H. Smith, J. Phys. C \textbf{5}, 2867 (1972).

\bibitem{Rendell81}
R. W. Rendell and S. M. Girvin, Phys. Rev. B \textbf{23}, 6610
(1981).

\bibitem{Novoselov04}
K. S. Novoselov \textit{et al.}, Science \textbf{306}, 666 (2004).

\bibitem{Dykhne94}
A. M. Dykhne and I. M. Ruzin, Phys. Rev. B \textbf{50}, 2369 (1994).

\bibitem{Burgess07}
C. P. Burgess and B. P. Dolan, Phys. Rev. B \textbf{76}, 113406
(2007).

\bibitem{Driscoll02}
T.~A.~Driscoll and L.~N.~Trefethen, {\it Schwarz-Christoffel
Mapping}, (Cambridge University Press, Cambridge, 2002).

\bibitem{ConformalMapping} Online Conformal Mapping dictionary, example 51:
http://math.fullerton.edu/mathews/c2003/ConformalMapDictionary.5.html


\bibitem{Giovannetti07}
G. Giovannetti \textit{et al.}, Phys. Rev. Lett. {\bf 101}, 026803
(2008).

\bibitem{Yacoby07}
J. Martin \textit{et al.}, Nat. Phys. {\bf 4}, 144 (2008).


\end{thebibliography}
\end{document}